\documentstyle[preprint,aps]{revtex}

\begin{document}
\draft
\title{Anomalous low-energy electron attachment in C$_{60}$}
\author{
\underline{Erio Tosatti}\thanks{Email: tosatti@tsmi19.sissa.it}}
\address{
International School for Advanced Studies (SISSA),
via Beirut 2-4, I-34013 Trieste, Italy.\\
International Centre for Theoretical Physics (ICTP),
P.O. BOX 586, I-34014 Trieste, Italy.}
\author{
Nicola Manini\thanks{Email: manini@tsmi19.sissa.it}}

\address{
International School for Advanced Studies (SISSA),
via Beirut 2-4, I-34013 Trieste, Italy.}

\date{Feb 26, 1994}
\maketitle
\begin{abstract}
We propose that thermal electron attachment to C$_{60}$ should occur
preferentially in the p-wave channel, following an analysis of the vibron
excitation spectrum of C$_{60}^-$. A very simple model based on this idea is
shown to account very well for recent attachment data. The unexplained
activation energy of $\approx$ 0.26 eV found experimentally is attributed to
the p-wave centrifugal barrier.
\end{abstract}


In a recent pair of papers, we have pointed out some remarkable properties of
fullerene molecular anions, C$_{60}^{n-}$, $n$=1,...5 \cite{assa1,assa2}. In
these ions, a threefold degenerate partly filled $t_{1u}$ molecular orbital
(MO) is linearly coupled to two $a_g$ and to eight $h_g$ molecular vibrations,
leading to a well-known Jahn-Teller (JT) effect
\cite{negri,sch,vzr,antropov,koga}. We have made two additional
observations, however.

The first is that a static JT description is not really adequate. Zero-point
effects are large in C$_{60}$ \cite{kohanoff}, while at the same time one
expects basically no barrier between equivalent JT potential minima, as
confirmed, e.g., by Hartree-Fock calculations \cite{koga}. A full quantum
treatment of ionic motion, typical of the {\em dynamical} JT effect is
therefore mandatory. As it turns out, fullerene anions constitute a linear
weak-coupling case which is easy to solve. In the end, one finds a remarkable
enhancement of the true stabilization energies relative to their classical
values, which makes the JT effect much more important than believed so far.

The second observation is that for odd $n$ and in particular for C$_{60}^-$,
the adiabatic coupling of $H_g$ ($L$=2) vibrational coordinates to the $t_{1u}$
($L$=1) electronic level \cite{comm0} is affected by a nonzero molecular Berry
phase \cite{mead}. Its presence, in turn, causes vibron quantization to be
anomalous, in qualitative analogy with, for instance Na$_{3}$ \cite{delac}. In
that molecule, a strong-coupling $e \otimes E$ dynamical JT case, the Berry
phase causes, very unusually, the total pseudorotational angular momentum
(excluding spin) to take up half-integer (instead of integer) values
\cite{delac}. In C$_{60}^-$, where ($t_{1u} \otimes H_g$) coupling is weak,
effects appear to be twofold. First, there is always an {\em odd}-$L$ state
which has energy lower than the even $L$ states in each multiplet of the
excitation ladder. In particular, the ground state is $L$=1, and the lowest
state on each $m$-vibron excited multiplet is $L$=2$m$+1. This is the analog
of the half-integer effect of Na$_{3}$. Secondly, $L$=0 states disappear
altogether from the low-lying excitation spectrum \cite{comm1}. For example,
each of the eight $H_g$ quadrupolar modes gives rise, after dynamical JT
coupling, to a one-vibron multiplet $L$=3,2,1 (in order of increasing energy),
followed by a two-vibron multiplet $L$=5,3,1,4,2,1,3 (in order of increasing
energy), etc., where in spherical geometry no $a_g$ ($L$=0) state appears.
Specifically, the absence of the $L$=0 states can be properly seen as a direct
consequence of the absence of $L$=1 states in the overtones of a quadrupolar
harmonic oscillator \cite{iachello}. Since, moreover, each of the two original
$A_g$ ($L$=0) modes gives rise to $t_{1u}$ ($L$=1) vibronic states only, while
all other remaining vibrations of different symmetry species \cite{negri} are
linearly uncoupled to the $t_{1u}$ electronic level of C$_{60}^-$, no other
$L$=0 state is produced. The C$_{60}$ molecule has in addition only a second
affinity level, of $t_{1g}$ symmetry, which lies roughly 1 eV above the
$t_{1u}$ \cite{haddon}, or 1.7 eV below the ionization level. This higher
electronic level is also coupled to molecular vibrations, giving rise to a
dynamical JT problem which is easily seen to be formally very similar to that
of the $t_{1u}$ level. In particular, therefore, this $t_{1g}$ level again
fails to give rise to $a_g$ ($L$=0) vibronic states in its spectrum
\cite{comm2}.

In this letter, we point out an unexpected consequence of the above scenario,
namely, an anomalous low-energy electron attachment to C$_{60}$. The absence of
$L$=0 bound states for C$_{60}^-$, as opposed to a whole set of vibronic $L$=1
states available down to the ground state chemical potential at $E_I$=-2.7 eV
\cite{yeretzian}, suggests that, unlike most common cases, such as, e.g.,
SF$_6$ \cite{christophorou}, electron attachment to C$_{60}$ should proceed by
{\em p-wave instead of s-wave scattering}. We show that this hypothesis allows,
via an extremely simple model, to account very well for the observed thermal
attachment rate, including its hitherto mysterious activated behaviour
\cite{smith}.

We model the free electron interaction with C$_{60}$ as a short-range
scattering event, each partial wave including elastic and inelastic channels.
If a channel with nonzero (inelastic) attachment cross section  must have bound
states in it, then we are led to assume that s-wave attachment can be
neglected, and must consider next the p-wave channel. Here, there will be a
centrifugal barrier $ \hbar^2/m_eR^2 $, whose traversal will constitute the
major obstacle to attachment. The estimated barrier height at an approximate
C$_{60}$ radius of $R\approx$ 10 a.u. is 10$^{-2}$ Hartrees or 0.27~eV,
encouragingly close to the observed and unexplained activation energy for
attachment \cite{smith}.

In order to make quantitative progress, we further model p-wave scattering and
attachment as follows. First, we assume that this inelastic process can be
replaced by an elastic p-wave scattering across the barrier, followed by an
instantaneous, infinitely efficient inelastic electron decay into the $t_{1u}$
bound state. In other words, we assume that once the p-wave electron has
crossed the centrifugal barrier it will be attached with probability 1. This
assumption seems justified by a very efficient distribution of energy among
many degrees of freedom of this large molecule \cite{comm3}. In
conclusion, the attachment cross section to be calculated under this assumption
coincides with the elastic p-wave cross section of a {\em real} potential
$V(r)$. In order to represent the effect of C$_{60}$ on a p-wave thermal
electron, this potential must be attractive (with at least a bound state) for
$r<R$ and vanish for $r>>R$.

As a first crude attempt we took for $V(r)$ a spherical square well of depth
$-V_0$ and radius $R$, with the constraint that is should have a single $L$=1
bound state at $E_I$=-2.7 eV. The p-wave phase shift $\delta_1$ for
this potential is \cite{schiff}
\begin{eqnarray}
	\delta_1&=&\tan^{-1}	\frac{j_1'(kR) - \gamma j_1(kR)}
				{n_1'(kR) - \gamma n_1(kR)} \ ,	\nonumber\\
	\gamma  &=&	\frac{ \chi j_1'(\chi R)}
		  	     { k    j_1(\chi R)}	    \ ,	\nonumber\\
	k  &=&\sqrt{ 2m_e E }/\hbar \ , \ \chi=\sqrt{ 2m_e (E+V_0) }/\hbar \ ,
\end{eqnarray}
where $j_1$ and $n_1$ are Bessel functions of order one. The corresponding
attachment cross section $\sigma_A(E)=12\pi \hbar^2(2m_eE)^{-1} \sin^2
\delta_1$ as a function of incident electron energy $E$ is shown in Fig.
\ref{sigma}. The cross section falls off as $E^2$ at small $E$ (due to barrier
tunneling) and as $E^{-2}$ at large $E$, with a maximum for $E$ of order of the
centrifugal barrier height and a zero due to the p-wave phase shift $\delta_1$
crossing $\pi$, as requested for one bound state. Since the barrier height is
considerable, it is clear that p-wave attachment will be very poor at thermal
energies. In s-wave, instead, the cross section would have been finite in the
$E\to 0$ limit, leading to the large thermal energy attachment normally
observed in other large molecules \cite{christophorou,lezius}.

{}From this cross section, the attachment rate for a thermal distribution with
temperature $T$
\begin{equation}
	A(T)= \frac{ 2^{1/2} } {m_e^{1/2} k_B T}
	\int_0^{\infty} {dE E^{1/2} \sigma_A(E) e^{-\frac{E}{k_B T}} }
\end{equation}
is easily calculated, and the result are plotted in Fig. \ref{fit}. The choice
of $R$ determines sensitively the correct order of magnitude for the absolute
attachment rate. The best fit is with $R$=5.27\AA, which is very reasonable for
C$_{60}$. The other parameter $V_0$ = 4.68 eV being fixed by the bound state
constraint, we find that the overall agreement with the thermal attachment
rates of Smith, \u{S}panel and M\"{a}rk is very good. The approximate activated
behaviour is retrieved, and related in our calculation to electrons thermically
negotiating the centrifugal barrier. Deviations in excess of a purely activated
behaviour seen both experimentally and theoretically at the lowest temperature
of 300 K, are found to be due to the $E^2$ quantum tunneling under the barrier.
At the highest temperature the calculated attachment rate is slightly smaller
than the experiment. This may be due to inadequacy of our simple $V(r)$ as well
as of our additional assumptions. The high-energy attachment cross section
\cite{lezius} shows strong peaks in the 1 - 8 eV region interpreted as Feshbach
resonances with electronic and vibrational states, which our simple model
clearly cannot account for. On the whole, however, we find that this extremely
crude model works well enough in the explored region of temperatures to make
further improvements unnecessary.

In conclusion, we have given arguments based on a recent dynamical JT study of
C$_{60}^-$, why low energy electron attachment to C$_{60}$ should occur
predominantly in the p-channel. We have furthermore shown that an extremely
simple model for that process yields a very good description of the
experimental attachment rate, including its anomalous activated behaviour,
which is now attributed to crossing of the centrifugal barrier. It is possible
that the phenomenon described here could also bear a connection to the
failure of the simple Dushman - Richardson formula to account quantitatively
for thermionic emission from C$_{60}^-$ ions \cite{yeretzian}. However, that
behaviour is shared by C$_{70}^-$ and C$_{96}^-$, whose physics is somewhat
different in many details. This problem requires a separate investigation.

\vskip 1cm

This work was supported by the Italian Istituto Nazionale di Fisica della
Materia INFM,the European US Army Research Office, the EEC under grants
ERBCHBGCT 920180, ERBCHRXCT 920062, ERBCHRXCT 930342, and NATO through CRG
920828.

\begin{figure}
\caption{
The $L$=1 cross section for three sets of values of the parameters $R$ and
$V_0$ that define the potential well $V(r)$. All choices of parameters satisfy
the constraint of a single bound p-state at $E_I$=-2.7 eV. The main peak
corresponds roughly to the centrifugal barrier height}
\label{sigma}
\end{figure}

\begin{figure}
\caption{
The electron attachment rate corresponding to three sets of values of the
parameters $R$ and $V_0$. The central curve is the best fit of the
experimental data points by Smith, \u{S}panel and M\"{a}rk
\protect\cite{smith}.}
\label{fit}
\end{figure}


\begin{references}
\bibitem{comm0} We shall use indifferently icosahedral or spherical symmetry
notations, at least when this causes no confusion.
\bibitem{comm1} We disregard here the true molecular rotations.
\bibitem{comm2}For $n$=0 or $n$=2 ($S^2$=0) instead, there is no Berry phase,
the ground state itself is $L$=0, and there is no such anomaly.
\bibitem{comm3}This is in contrast with the s-wave channel, where there is no
barrier, but no negative energy states available for fast attachment.
\end{references}
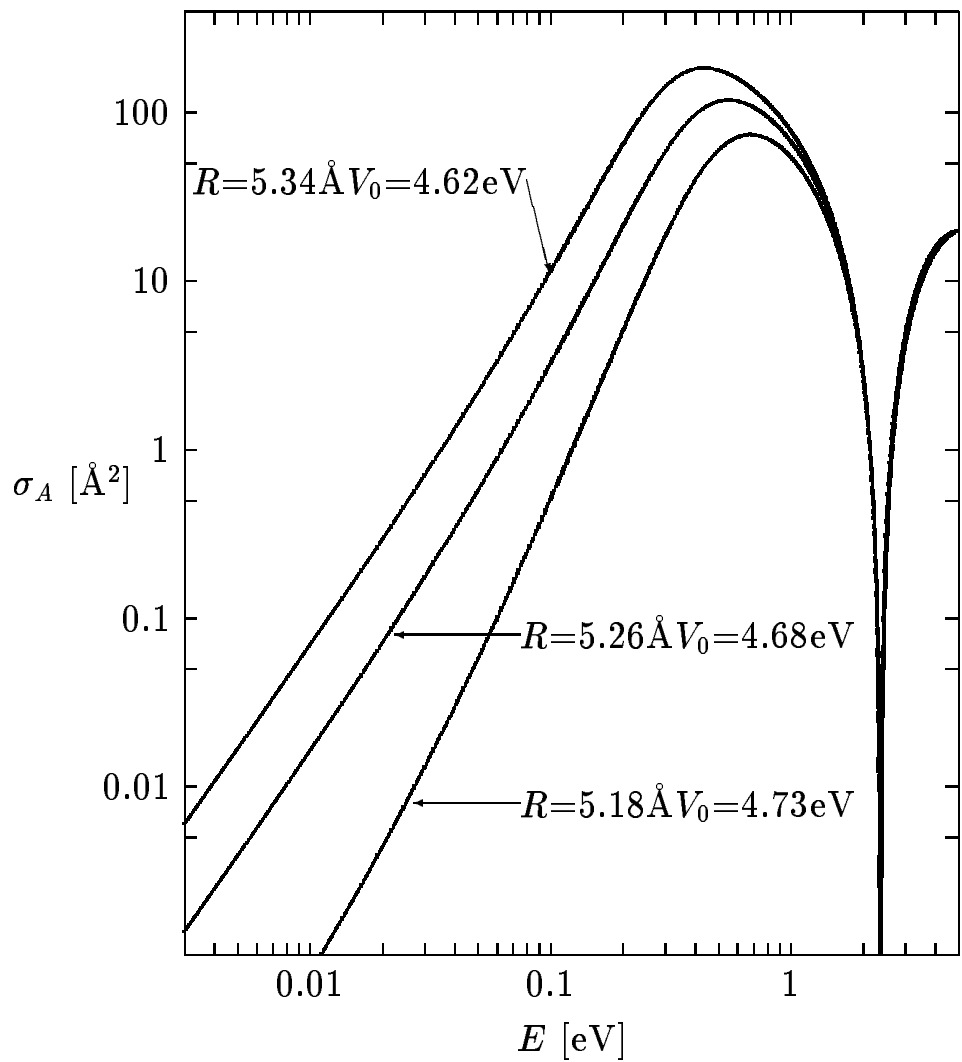
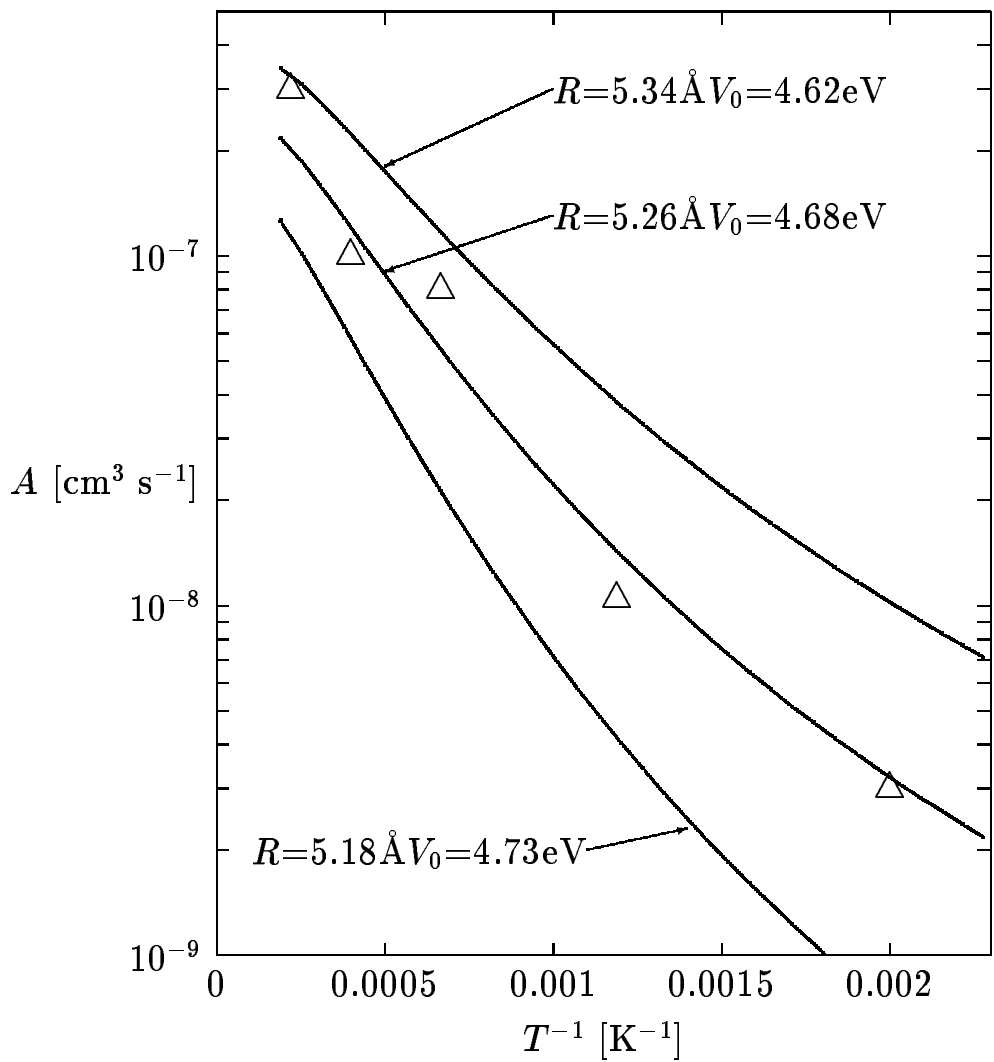
\end{document}